\documentclass{article}
\usepackage[preprint]{neurips_2023}

\usepackage[utf8]{inputenc} 
\usepackage[T1]{fontenc}    
\usepackage{hyperref}       
\usepackage{url}            
\usepackage{booktabs}       
\usepackage{amsfonts}       
\usepackage{nicefrac}       
\usepackage{microtype}      
\usepackage{xcolor}         
\usepackage{multirow}
\usepackage{graphicx}

\title{Learnable Gabor kernels in convolutional neural networks for seismic interpretation tasks}

\author{Fu Wang, Tariq Alkhalifah\\
King Abdullah University of Science and Technology\\
\{fu.wang, tariq.alkhalifah\}@kaust.edu.sa.}

\begin{document}

\maketitle

\begin{abstract}
The use of convolutional neural networks (CNNs) in seismic interpretation tasks, like facies classification, has garnered a lot of attention for its high accuracy. However, its drawback is usually poor generalization when trained with limited training data pairs, especially for noisy data. Seismic images are dominated by diverse wavelet textures corresponding to seismic facies with various petrophysical parameters, which can be suitably represented by Gabor functions.
Inspired by this fact, we propose using learnable Gabor convolutional kernels in the first layer of a CNN network to improve its generalization. 
The modified network combines the interpretability features of Gabor filters and the reliable learning ability of original CNN. More importantly, it replaces the pixel nature of conventional CNN filters with a constrained function form that depends on 5 parameters that are more in line with seismic signatures.
Further, we constrain the angle and wavelength of the Gabor kernels to certain ranges in the training process based on what we expect in the seismic images. The experiments on the Netherland F3 dataset show the effectiveness of the proposed method in a seismic facies classification task, especially when applied to testing data with lower signal-to-noise ratios.
Besides, we also test this modified CNN using different kernels on salt$\&$pepper and speckle noise. The results show that we obtain the best generalization and robustness of the CNN to noise when Gabor kernels are used in the first layer.

\end{abstract}

\section{Introduction}

Seismic interpretation is a fundamental step in the seismic exploration value chain, and it is often the most critical step in the decision-making process \citep{dumay1988multivariate}. Within the seismic interpretation tasks, identifying seismic facies from seismic images plays a vital role in hydrocarbon exploration and development. Seismic facies with different petrophysical parameters often induce different seismic responses. The identification of seismic facies can be regarded as a pattern-recognition problem. The traditional manual interpretation of seismic facies is highly dependent on the skills of interpreters and it is also time-consuming. In order to overcome these limitations, multiple seismic attributes with explainable correspondence have been proposed to assist seismic interpretation. Subsequently,  some machine learning methods including support vector machine \citep{zhao2014lithofacies,zhang2015brittleness,wrona2018seismic}, the random forest \citep{ao2019identifying}, self-organizing maps \citep{saraswat2012artificial,zhao2017constraining}, generative topographic mapping \citep{roy2014generative}, and independent component analysis \citep{lubo2019independent} use these attributes to classify seismic facies.

Recently, with the rising popularity of deep learning, many have proposed using convolutional neural networks (CNNs) to promote seismic interpretation tasks with reasonable success, like seismic facies classification \citep{zhao2018seismic,  di2020seismic,feng2021bayesian,liu2020seismic}, salt body identification \citep{waldeland2017salt, shi2019saltseg, di2020comparison}, fault detection \citep{di2018seismic,zhao2018fault, wu2019faultseg3d,wu2019faultnet3d}, and horizon tracking \citep{wu2019semiautomated, tschannen2020extracting}. The advantage of CNNs is their strong representation and nonlinear mapping abilities in handling large datasets, which improve the speed and accuracy of seismic facies interpretation. However, CNNs also have some disadvantages. Without enough training data pairs, they can easily overfit, resulting in poor generalization. In other words, the performance on tests and especially inference data could be poor, which is a serious limitation of CNNs. One of the many suitable ways to remedy this weakness is to embed prior information into CNNs. Specifically, for seismic facies classification, the input seismic images mainly include texture features constituted by band-limited wavelets and some additive or correlated noise. Thus, we consider the Gabor filter as a good prior to constraining the CNNs. Seismic data have long been represented and filtered by Gabor functions \citep{womack1994seismic}. For machine learning, \cite{daugman1985uncertainty} first proposed the Gabor function to model the spatial summation properties (of the receptive fields) of simple cells in the visual cortex. Soon afterward, Gabor filters were widely used in diverse pattern analysis applications and are regarded as an efficient tool in extracting spatial local texture features. Besides, \cite{krizhevsky2017imagenet} demonstrated that after training on real-life images, the first convolution layer of deep CNNs tends to learn Gabor-like filters. Thus, using Gabor filters in the first layer of a CNN for seismic pattern recognition problems could provide more reliable results.

In this work, encouraged by the good interpretability of the Gabor filter and the reliable learning ability of CNNs, we use a Gabor convolution kernel in the first layer of a CNN for seismic facies classification, as an example of interpretation task. Such a modification will replace the pixel focus of the common CNN filters with a constrained function form that is in line with seismic signatures. As a result, the Gabor filter parameters are learnable, and these parameters, like wavelength and angle, provide identifiable characteristics of seismic signals. In fact, each Gabor kernel is controlled by 5 parameters with distinct meanings and can be constrained within the range we expect in the seismic images. Thus, we utilize constraints on the wavelength and angle range of the Gabor filter.  We test our method on the Netherland F3 datasets to validate the features gained by replacing the conventional convolution kernel with the Gabor kernel.  

Specifically, the main contributions of this paper can be summarized as follows:
\begin{itemize}
    \item We propose a simple modified CNN with Gabor kernels in the first layer, which embeds the prior seismic texture into neural networks.
    We regularize certain parameters of Gabor kernels to filter the input based on our expectations of seismic images.
    \item Our approach provides new insights into incorporating the prior information provided by Gabor functions into CNN for seismic facies classification, and these insights could be easily adapted to other seismic data tasks. 
    \item We evaluate the effectiveness of the CNN with Gabor kernels on seismic images with lower signal-noise ratios, and with added salt$\&$pepper and speckle noise, and show the improvements in accuracy compared to conventional CNN.
\end{itemize}

The rest of the paper is organized as follows: we first introduce the Gabor function. Next, we explain how to replace Conv kernels with Gabor kernels in the first layer of CNN, and how to regularize the parameter of Gabor kernels during the training. Then we show the results of conventional CNN and our modified CNN, and the modified CNN with constrained Gabor kernels. To further validate the generalization and robustness to noise, we also test on salt$\&$pepper and speckle noise. Finally, we share out thoughts and summarize our developments in the discussion and conclusion sections.

\section{Method}

\subsection{The Gabor function}
A two-dimensional Gabor function is defined as a complex sinusoidal plane wave weighted by a Gaussian filter, as follows
\begin{equation}
g(x, y ; \lambda, \theta, \phi, \sigma, \gamma)=\exp \left(-\frac{x^{\prime 2}+\gamma^2 y^{\prime 2}}{2 \sigma^2}\right) \exp \left(i\left(2 \pi \frac{x^{\prime}}{\lambda}+\phi\right)\right),
\label{equ1}
\end{equation}
where $x^{\prime}=x \cos \theta+y \sin \theta$ and $y^{\prime}=-x \sin \theta+y \cos \theta$; $i$ is the imaginary unit and $\Theta=\{\lambda, \theta, \phi, \gamma, \sigma\}$ are the parameters controlling the shape of Gabor function. 
Thereinto, $\lambda$ represents the wavelength of the sinusoidal plane wave component of the Gabor function; $\theta$ defines the orientation (angle) of the plane wave;  $\phi$ is the phase shift; $\gamma$ represents the ellipticity of the Gaussian support of the Gabor function; $\sigma$ controls the standard deviation of the Gaussian filter of the Gabor function.

\begin{figure*}[!htb]
  \centering
  \includegraphics[width=1\columnwidth]{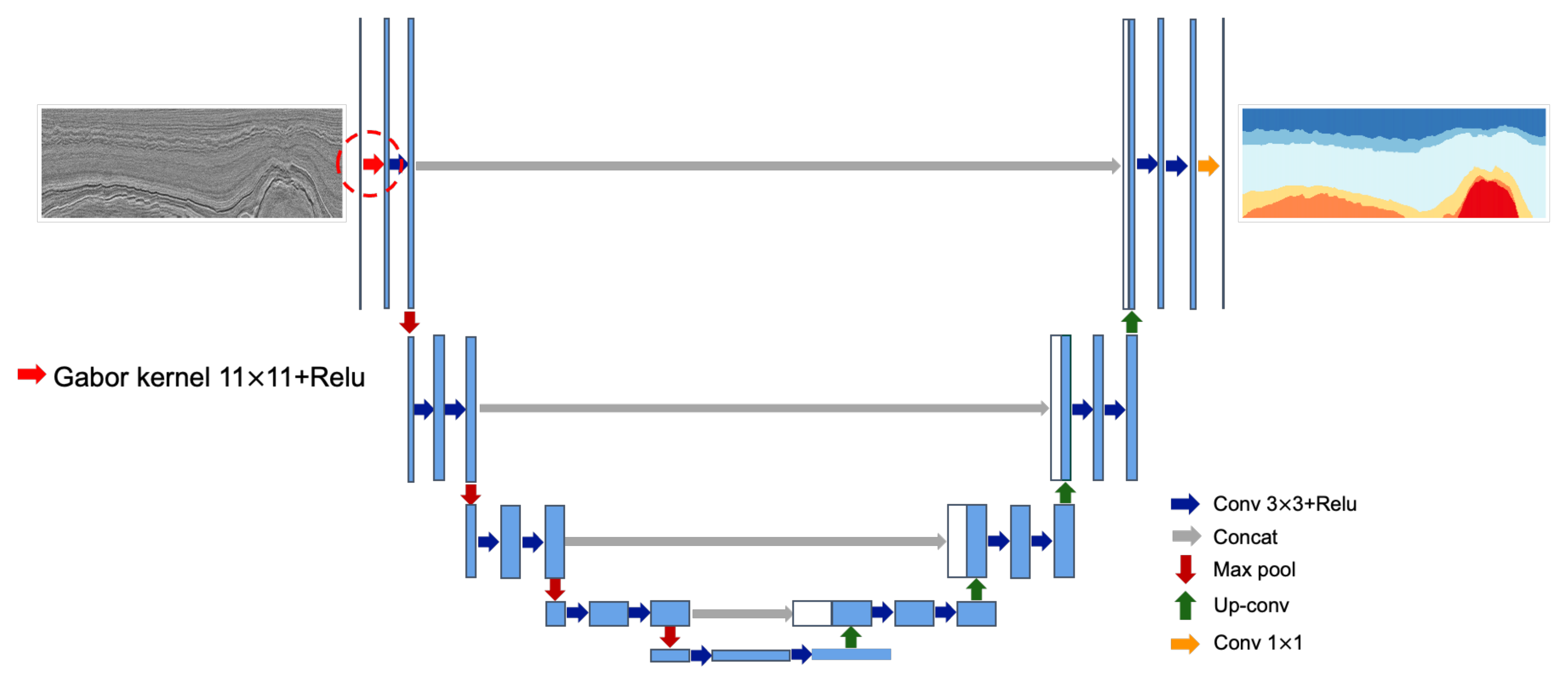}
  \vspace{-12pt}
  \caption{The U-Net structure with learnable Gabor kernels in the first layer.}
  \vspace{-6pt}
\label{network}
\end{figure*}

Equation \ref{equ1} can be rewritten in real numbers form, by splitting it into its real and imaginary parts, with the real part
\begin{equation}
g_r(x, y ; \lambda, \theta, \phi, \sigma, \gamma)=\exp \left(-\frac{x^2+\gamma^2 y^{\prime 2}}{2 \sigma^2}\right) \cos \left(2 \pi \frac{x^{\prime}}{\lambda}+\phi\right), 
\label{equ2}
\end{equation}
and the imaginary part
\begin{equation}
g_i(x, y ; \lambda, \theta, \phi, \sigma, \gamma)=\exp \left(-\frac{x^{\prime 2}+\gamma^2 y^2}{2 \sigma^2}\right) \sin \left(2 \pi \frac{x^{\prime}}{\lambda}+\phi\right),    
\end{equation}
In this work, we use (equation \ref{equ2}) to replace the classic convolution kernel in the first layer of our CNN, as shown in Figure \ref{network}.

\subsection{Learnable Gabor kernels in CNNs}
Traditionally, we specify the Gabor function parameters in advance before we use them as a filter. Gabor filters generated in this way may not be optimal for certain inputs. Besides, it is not an easy task to select suitable parameter values for the Gabor function. So in our work, all the parameters in the Gabor function are potentially learned from the training data in the training process. In other words, we use Gabor convolution kernels that adaptively update their parameters instead of using hand-crafted parameters.
Compared with classical convolution kernels in which each pixel is learned independently, the parameters $\Theta=\{\lambda, \theta, \phi, \gamma, \sigma\}$ in the Gabor kernel constrain the relation between the pixels to produce the Gabor filter.
In addition, a Gabor convolution kernel only has 5 parameters, whereas a classical convolution with kernel size $k$ will have $k^2$ independent pixels to determine. Luckily, the Gabor convolution module can be embedded into any network and we can train the 5 parameters using the gradient descent method as part of the full network for as many CNN channels we need to represent the input. 

\subsection{Constrained Gabor kernels in first layers}
Due to the distinct meaning of the five parameters in the Gabor convolution kernel, we can regularize them to control (filter) the input. For seismic images, two key parameters are the wavelength and direction of the plane wave. In this work, we set a range for these parameters that is in line with what we want to include from the seismic images for facies classification. This range can also help reduce the propagation of noise into our CNN.

\section{Numerical examples}
We use part of the Netherland F3 dataset (specifically, inlines [300,700] and crosslines [300,1000]), which includes six types of facies \citep{alaudah2019machine} to demonstrate the generalization abilities of our proposed method. The dataset includes six groups of lithostratigraphic units, from top to bottom consisting of the Upper North Sea group, the Middle North Sea group, the Lower North Sea group, the Rijnland/chalk group, the Scruff group, and the Zechstein group. We uniformly select 21 of the 401 inline profiles as the training dataset, while testing on the rest. To make it more challenging, we add 0.4dB random noise to the training dataset, but add random noise with the same and lower signal-to-noise ratios to the testing dataset. 
The backbone network used here is UNet with 23 layers.
First, we test two conventional UNets with 3$\times$3 and 11$\times$11 Conv Kernels, and a modified UNet with a Gabor convolutional kernels (11$\times$11) without any constraints in the first layer. As for the residual layers, we use 3$\times$3 Conv Kernels.
Then we test the Gabor kernels with different angles constraints ($\theta\in [-\pi/6,\pi/6]$, $[-\pi/4,\pi/4]$ and $[-\pi/3,\pi/3]$) and wavelength constraints ($\lambda \in [4, +\infty)$, $[8, +\infty)$, $[12, +\infty)$), respectively. 
Besides, to further test the generalization of the Gabor kernels, we include salt$\&$pepper and speckle noise in our tests.

Here we use the following metrics: pixel accuracy (PA), class accuracy (CA) for a single class, mean class accuracy (MCA) for all the classes, mean class accuracy (MCA), frequency-weighted intersection over union (FWIU), and mean intersection over union (Mean IU) to evaluate the performance of seismic facies segmentation with various kernels in the first layer. A detailed explanation of these metrics is provided in Appendix.
\subsection{The comparison between Gabor kernel and Conv kernel}
\begin{figure*}[!htb]
  \centering
  \includegraphics[width=1\columnwidth]{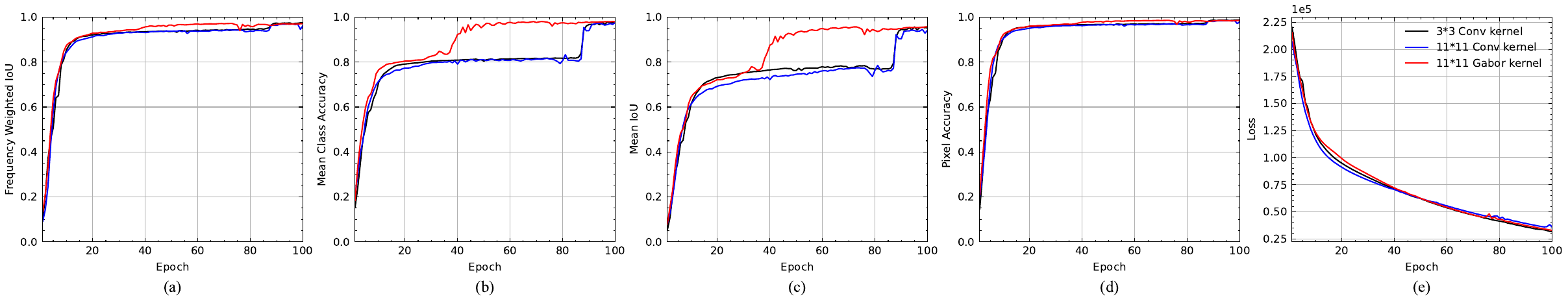}
  \vspace{-12pt}
  \caption{The training metric curves for 21 inline profiles uniformly chosen from the whole 401 inline profile with 0.4dB Gaussian noise, including (a)FWIU, (b)MCA, (c)Mean IU, (d)PA, and (e)loss, for the 3$\times$3 and 11$\times$11 Conv kernels and the 11$\times$11 Gabor kernels.}
  \vspace{-6pt}
\label{conv3conv11gabor1102train}
\end{figure*}

\begin{figure*}[!htb]
  \centering
  \includegraphics[width=1\columnwidth]{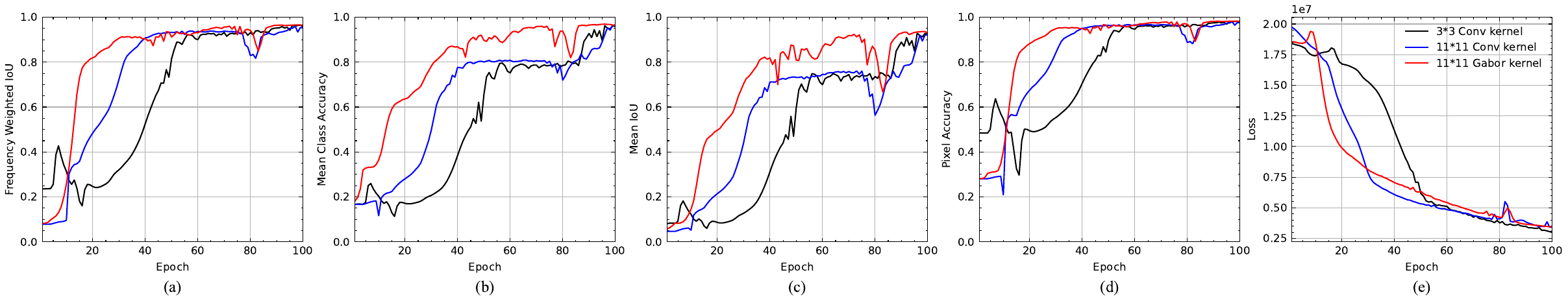}
  \vspace{-12pt}
  \caption{The testing metric curves for the rest (validation) of the 380 inline profiles with 0.4dB white Gaussian noise, including (a)FWIU, (b)MCA, (c)Mean IU, (d)PA, and (e)loss, for the 3$\times$3 and 11$\times$11 Conv kernels and the 11$\times$11 Gabor kernels.}
  \vspace{-6pt}
\label{conv3conv11gabor1102test}
\end{figure*}

We first display the training metric curves, including FWIU, MCA, Mean IU, PA, and loss, for the 3$\times$3 and 11$\times$11 Conv kernels and the 11$\times$11 Gabor kernels in Figure \ref{conv3conv11gabor1102train}. 
Figure \ref{conv3conv11gabor1102test} shows the corresponding test metric curves for the test data with 0.4dB random noise. 
During the training, we observe that the FWIU and PA of the CNNs with 3$\times$3 and 11$\times$11 Conv kernels or the 11$\times$11 Gabor kernel are close to each other, though there is an improvement using the Gabor kernel.
While in terms of MCA and Mean IU, the CNN with the 11$\times$11 Gabor kernel has noticeable improvements compared to the classic conventional CNN, where the CNN with the Gabor kernel converges much faster (at 45 epochs) than those with the Conv kernels, which converge at 90 epochs.
Recall that FWIU and PA (see Appendix), unlike the other measure,  reflect the overall performance of the classification. Still, the MCA and Mean IU provide average scores for different classes, which means the high score only happens when the accuracy for each class reaches a high level.
Thus, metrics demonstrate that the CNN with the Gabor kernel could quickly fit the classes with fewer samples.

At the testing stage, we observe that when the training accuracy reaches the same level, the testing accuracy of the CNN with the Gabor kernel is much better than that of the CNN with Conv kernel. This implies that the CNN with the Gabor kernel has better generalization properties.

\begin{figure*}[!htb]
  \centering
  \includegraphics[width=1\columnwidth]{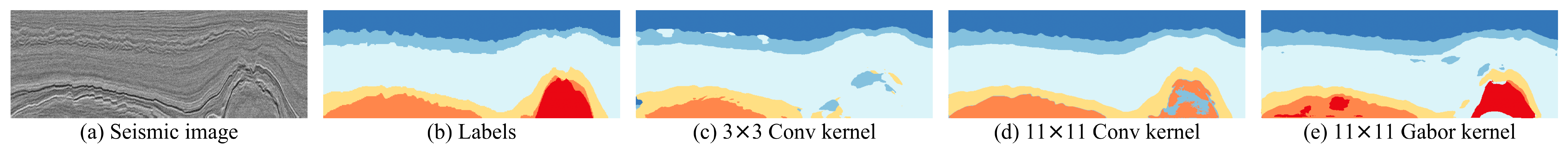}
  \vspace{-12pt}
  \caption{The prediction results for inline \#355 with 0.4dB Gaussian noise using the trained CNN models after 50 training epochs in Figure \ref{conv3conv11gabor1102train}. }
  \vspace{-6pt}
\label{conv3conv11gabor1102test50epochpred55}
\end{figure*}

\begin{figure*}[!htb]
  \centering
  \includegraphics[width=1\columnwidth]{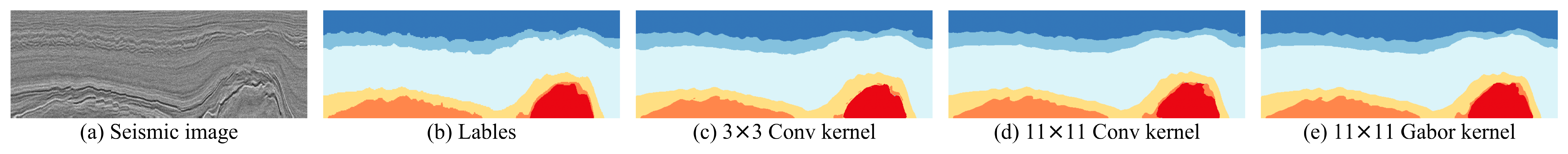}
  \vspace{-12pt}
  \caption{The prediction results for inline \#375 with 0.4dB Gaussian noise using the trained CNN models after 50 training epochs in Figure \ref{conv3conv11gabor1102train}. }
  \vspace{-6pt}
\label{conv3conv11gabor1102test100epochpred75}
\end{figure*}

Figure~\ref{conv3conv11gabor1102test50epochpred55} is the visualization of the predictions for inline \#355 using the trained model after 50 training epochs. 
It shows that with limited training epochs, the CNN with Conv kernel could not reasonably predict the classes with limited training samples, while the CNN with Gabor kernel could achieve that.
Figure \ref{conv3conv11gabor1102test100epochpred75} is the visualization of the predictions for inline \#375 using the trained model after 100 training epochs. 
For this case, all types of CNN predict well. 

\begin{figure*}[!htb]
  \centering
  \includegraphics[width=1\columnwidth]{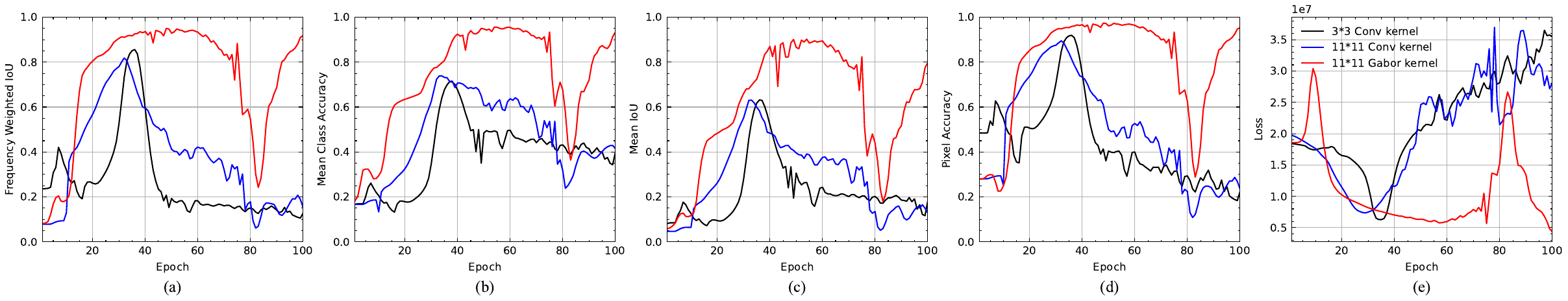}
  \vspace{-12pt}
  \caption{The testing metric curves for the rest (validation) of the 380 inline profiles with -5.6 dB Gaussian noise, including (a)FWIU, (b)MCA, (c)Mean IU, (d)PA, and (e)loss, for the 3$\times$3 and 11$\times$11 Conv kernels and the 11$\times$11 Gabor kernels.}
  \vspace{-6pt}
\label{conv3conv11gabor1104test}
\end{figure*}

To further test the generalization of the CNN with different kernels, we test these types of CNNs on testing images containing -5.6 dB white Gaussian noise, shown in Figure~\ref{conv3conv11gabor1104test}.
We see that even when dealing with noisy images, whose signal-to-noise level is lower than that of the training data, the Gabor kernels still produce good performance when considering early stopping.
It means that the CNN with the Gabor kernel is more robust to the noise.

\begin{figure*}[!htb]
  \centering
  \includegraphics[width=1\columnwidth]{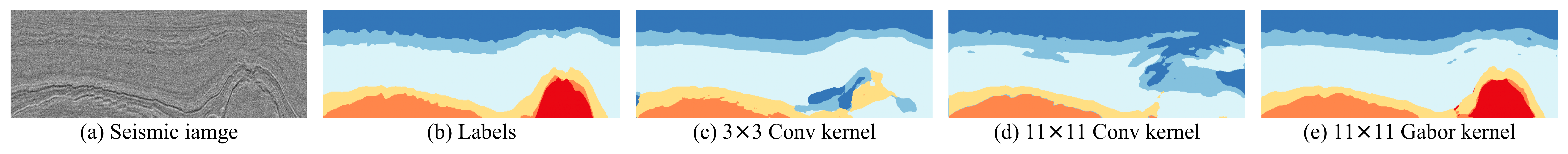}
  \vspace{-12pt}
  \caption{The prediction results for inline \#355 with -5.6 dB Gaussian noise using the trained CNN models  with different epochs, for 3$\times$3 Conv kernel is 36 epoch, for 11$\times$11 Conv kernel is 32 epoch, and for 11$\times$11 Gabor kernel is 50 epoch.}
  \vspace{-6pt}
\label{conv3conv11gabor1104test50epochpred55}
\end{figure*}

\begin{figure*}[!htb]
  \centering
  \includegraphics[width=1\columnwidth]{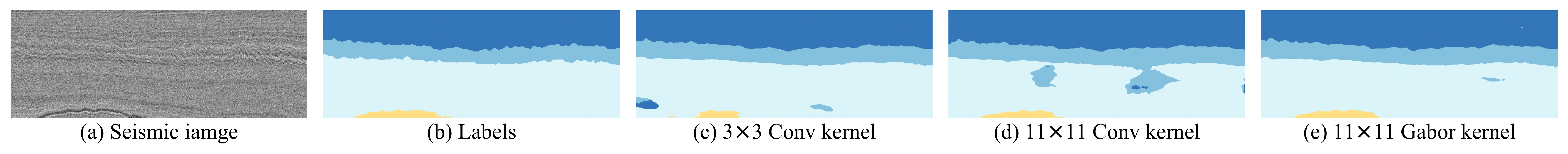}
  \vspace{-12pt}
  \caption{The prediction results for inline \#609 with -5.6 dB Gaussian noise using the trained CNN models with different epochs, for 3$\times$3 Conv kernel is 36 epoch, for 11$\times$11 Conv kernel is 32 epoch, and for 11$\times$11 Gabor kernel is 50 epoch.}
  \vspace{-6pt}
\label{conv3conv11gabor1104test50epochpred309}
\end{figure*}

In this case, we adopt the early stopping and evaluate the CNNs. 
We show the predictions for inlines \#355 and \#609 in Figures~\ref{conv3conv11gabor1104test50epochpred55} and \ref{conv3conv11gabor1104test50epochpred309}, respectively.
The CNN with 3$\times$3 Conv kernel is the model trained for 36 epochs, the CNN with 11$\times$11 Conv kernel is trained for 32 epochs, and the CNN with 11$\times$11 Gabor kernel is trained for 50 epochs.
We can again see that the CNN with the Gabor kernel is robust to noise.

\subsection{Constraining the Gabor parameters}
\begin{figure*}[!htb]
  \centering
  \includegraphics[width=1\columnwidth]{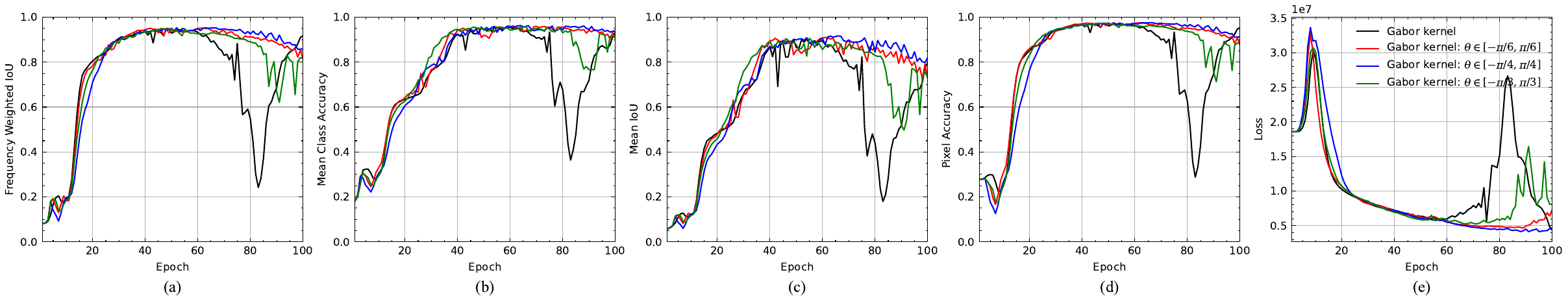}
  \vspace{-12pt}
  \caption{The testing metric curves of the CNN with the Gabor kernels under different constraints of angles (no-constraints, $\theta\in [-\pi/6,\pi/6]$, $[-\pi/4,\pi/4]$ and $[-\pi/3,\pi/3]$) using the testing data containing -5.6 dB Gaussian noise.}
  \vspace{-6pt}
\label{gabor11gabor11w30gabor11w45gabor11w6004test}
\end{figure*}

\begin{figure}[!htb]
  \centering
  \includegraphics[width=0.5\columnwidth]{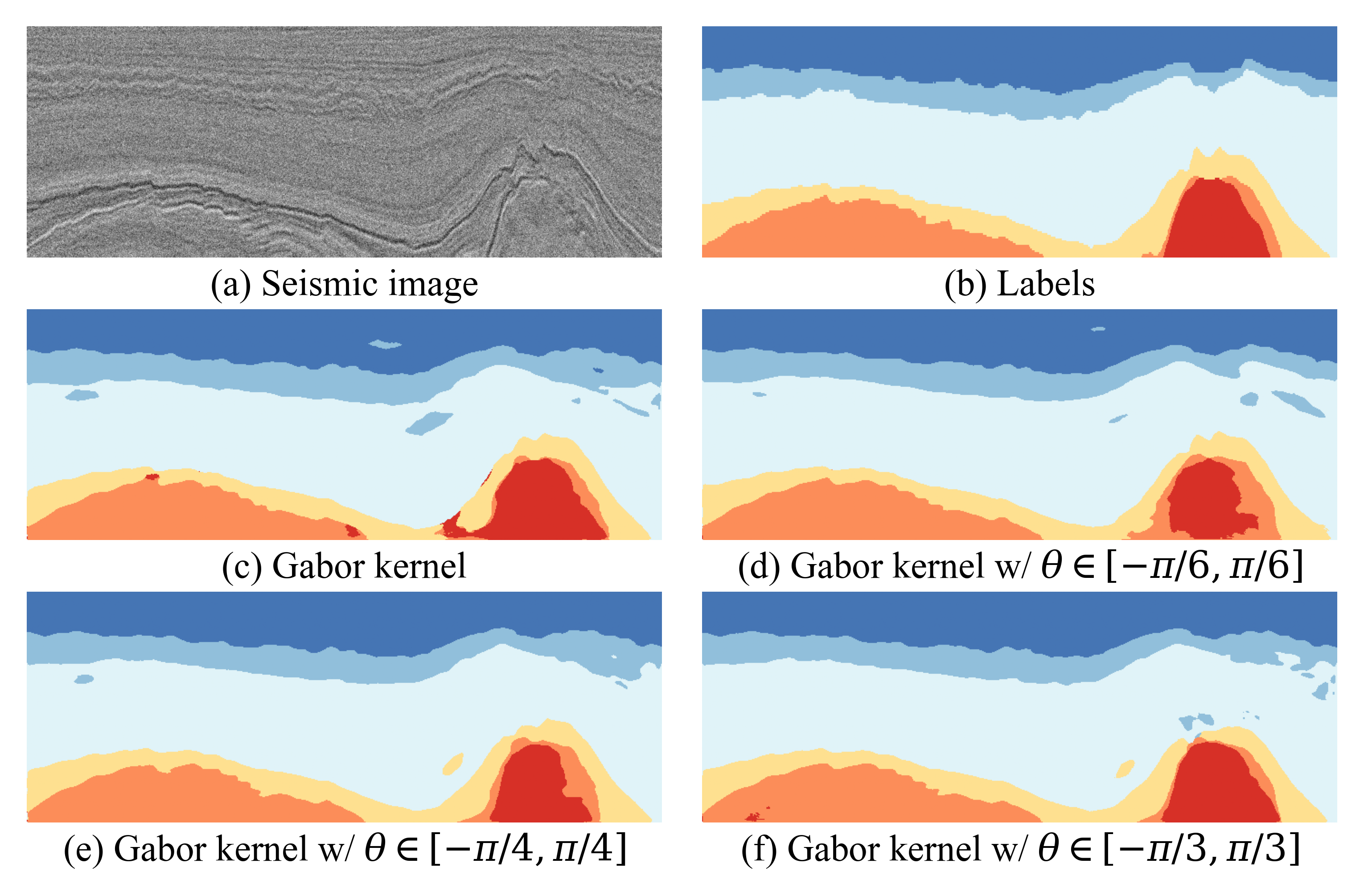}
  \vspace{-12pt}
  \caption{The prediction results for inline \#333 with -5.6 dB Gaussian noise using the trained CNN models with different angle constraints in the Gabor kernel after 50 epochs.}
  \vspace{-6pt}
\label{gaborgabor30gabor45gabor6004pred33}
\end{figure}

As we know, the most obvious and important features in seismic imaging results, for interpretation, are textures, angles, and frequencies.
As mentioned before, the beauty of the Gabor kernels is that we could control these parameters, as they define the Gabor basis function, to improve the performance of the CNN with Gabor kernels and also to avoid overfitting to some extent.
Thus, in this section, we focus on tests where parameters of the Gabor kernel in CNN are constrained within a range.
Specifically, we apply constraints on the angles and wavelengths of the Gabor basis function.
Figure~\ref{gabor11gabor11w30gabor11w45gabor11w6004test} shows the training metric curves of the CNN with the Gabor kernels under different constraints of angles using the training data containing 0.4 dB white Gaussian noise.
The constraints admit slightly faster convergences than that without the constraints.
Figure~\ref{gaborgabor30gabor45gabor6004pred33} is the corresponding test curves on the testing data containing -5.6 dB Gaussian noise. 
It is obvious that  with the angle constraint, the CNN with Gabor kernels is much better than that without the constraint, avoiding overfitting to some extent.
We further show in Figure~\ref{gaborgabor30gabor45gabor6004pred33} the predictions on the inline \#333 after 50 training epochs .
When we constrain the angle $\theta\in[-\pi/6, \pi/6]$, for the facies denoted by the red color, the CNN with the Gabor kernel has a hard time identifying the class where the slope of the texture is large. 
As we increase the range of the angles in the Gabor kernel, the CNN could identify the red class with higher accuracy.
\begin{figure*}[!htb]
  \centering
  \includegraphics[width=1\columnwidth]{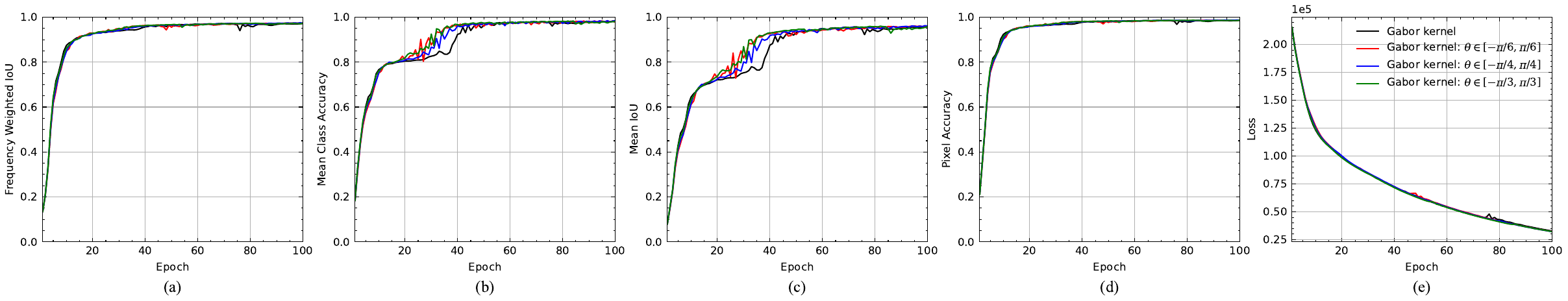}
  \vspace{-12pt}
  \caption{The training metric curves of the CNN with the Gabor kernels under different constraints of angles (no-constraints, $\theta\in [-\pi/6,\pi/6]$, $[-\pi/4,\pi/4]$ and $[-\pi/3,\pi/3]$) using the training data containing 0.4 dB  Gaussian noise.}
  \vspace{-6pt}
\label{gabor11gabor11w30gabor11w45gabor11w6002train}
\end{figure*}

\begin{figure*}[!htb]
  \centering
  \includegraphics[width=1\columnwidth]{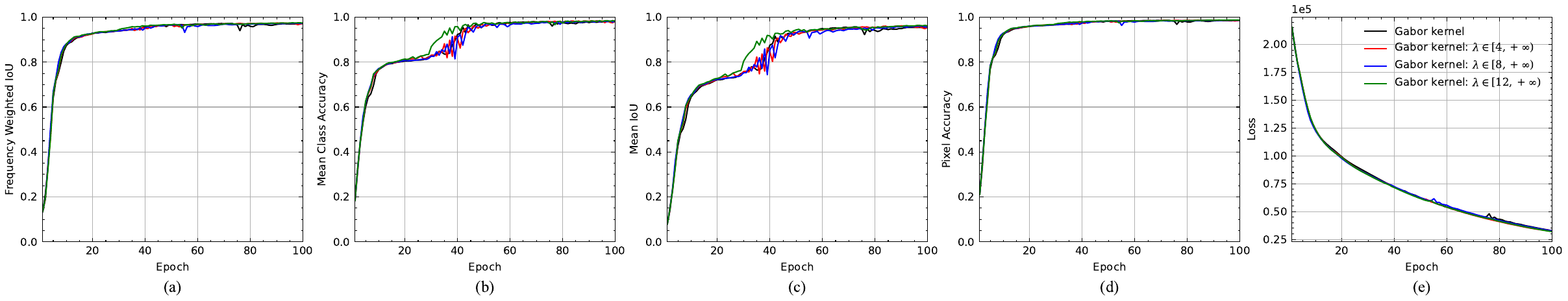}
  \vspace{-12pt}
  \caption{The training metric curves of the CNN with the Gabor kernels under different constraints of wavelengths (no-constraints, $\lambda \in [4, +\infty)$, $[8, +\infty)$, $[12, +\infty)$) using the training data containing 0.4 dB Gaussian noise.}
  \vspace{-6pt}
\label{gabor11gabor11w4gabor11w6gabor11w8gabor11w10gabor11w1202train}
\end{figure*}

\begin{figure*}[!htb]
  \centering
  \includegraphics[width=1\columnwidth]{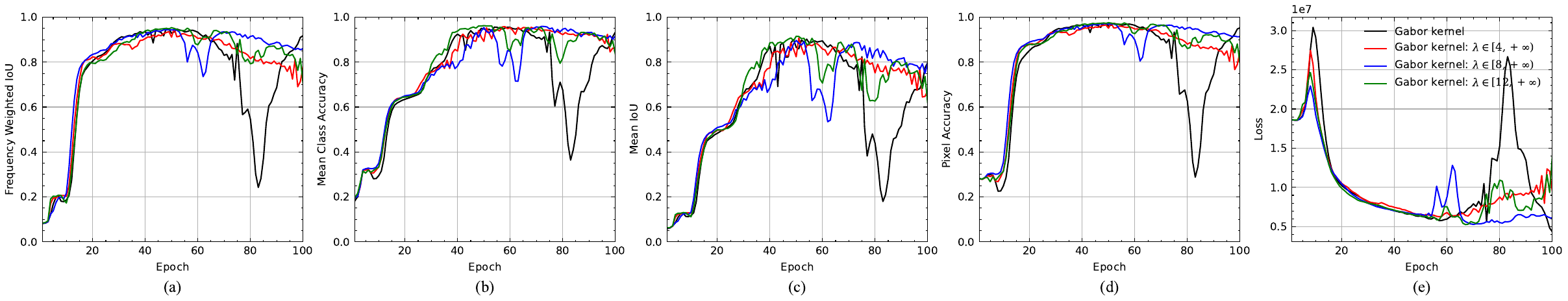}
  \vspace{-12pt}
  \caption{The testing metric curves of the CNN with the Gabor kernels under different constraints of wavelengths (no-constraints, $\lambda \in [4, +\infty)$, $[8, +\infty)$, $[12, +\infty)$) using the testing data containing -5.6 dB Gaussian noise.}
  \vspace{-6pt}
\label{gabor11gabor11w4gabor11w8gabor11w1204test}
\end{figure*}

\begin{figure}[!htb]
  \centering
  \includegraphics[width=0.5\columnwidth]{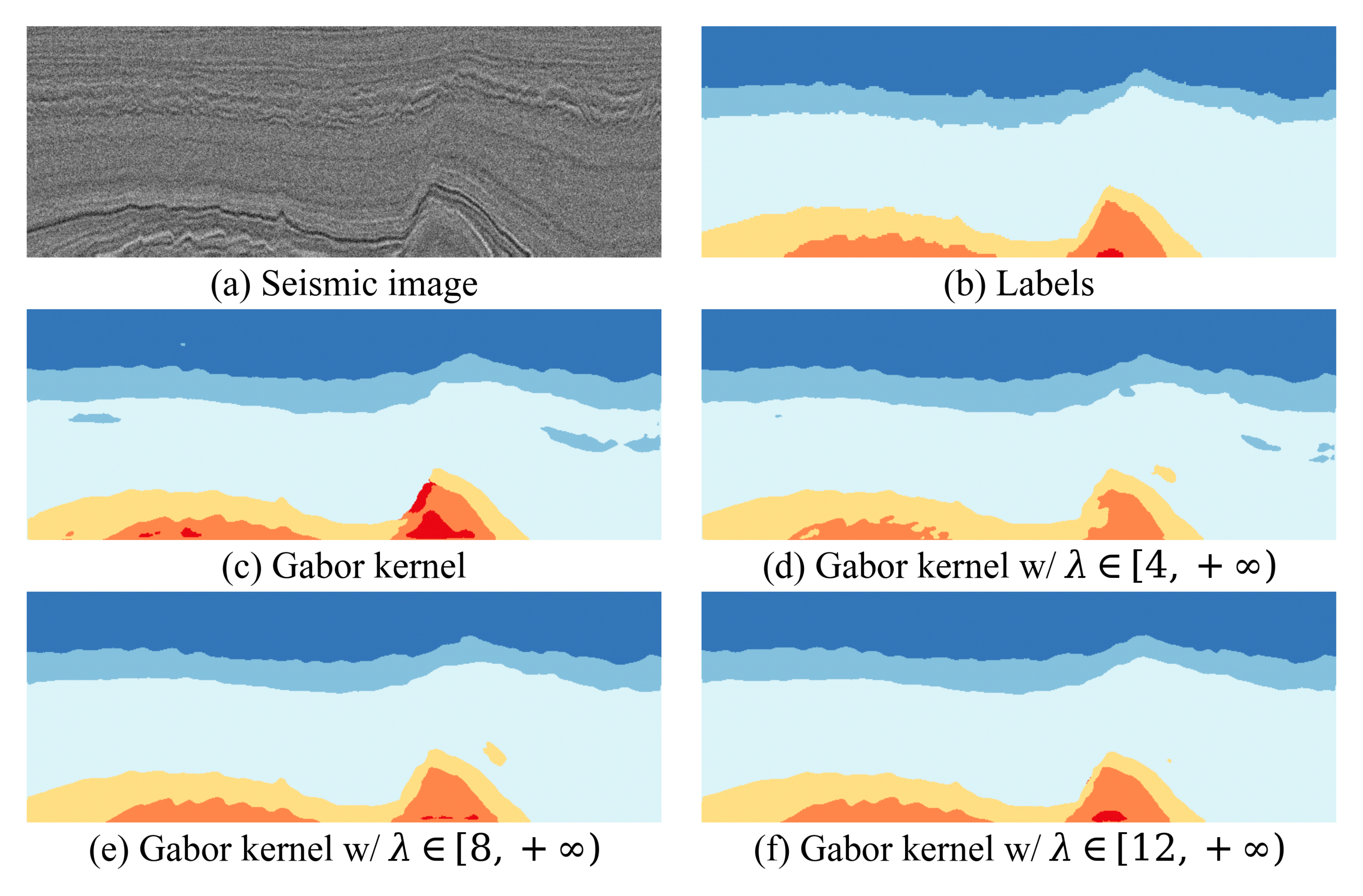}
  \vspace{-12pt}
  \caption{The prediction results for inline \#465 with -5.6 dB white Gaussian noise using the trained CNN models with different wavelength constraints in the Gabor kernel after 50 epochs.}
  \vspace{-6pt}
\label{gaborgabor4gabor8gabor1204pred165}
\end{figure}

Figures~\ref{gabor11gabor11w4gabor11w6gabor11w8gabor11w10gabor11w1202train} shows the training metric curves for the CNN with the Gabor kernels under different constraints applied to the wavelength using the training data containing 0.4 dB Gaussian noise.
The CNN with a $[12,\infty]$ wavelength constraint, converges slightly faster.
Figure~\ref{gabor11gabor11w4gabor11w8gabor11w1204test} shows the corresponding test metric curves on the data containing -5.6 dB Gaussian noise. 
It is obvious that with the wavelength constraint, the CNN with Gabor kernels is more stable than that without the constraint and also avoids overfitting to some extent.
We further show in Figure~\ref{gaborgabor4gabor8gabor1204pred165} the predictions on the inline \#465 after 50 training epochs.
When the minimum wavelength of the Gabor kernel increases, the predictions for the classes, where the imaging result contains long-wavelength components, are improving.
The application of constraints on particular Gabor parameters helps the training of the other Gabor parameters, as we reduce the search space.
So, the improvements of the CNN with a constrained Gabor kernel are not only the contribution of the constrained parameters but also the other parameters.

\subsection{Test on salt$\&$pepper and speckle noises}
\begin{figure*}[!htb]
  \centering
  \includegraphics[width=1\columnwidth]{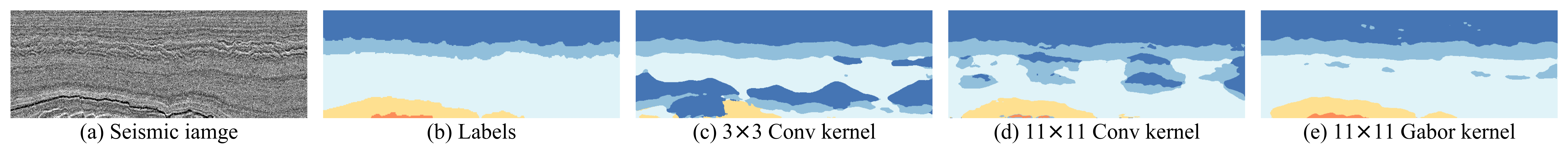}
  \vspace{-12pt}
  \caption{The prediction results for inline \#510 with 30\% salt $\&$ pepper noise using the same CNN models as that in Figures~ \ref{conv3conv11gabor1104test50epochpred55} and \ref{conv3conv11gabor1104test50epochpred309} for different kernels.}
  \vspace{-6pt}
\label{saltpepper210}
\end{figure*}

\begin{figure*}[!htb]
  \centering
  \includegraphics[width=1\columnwidth]{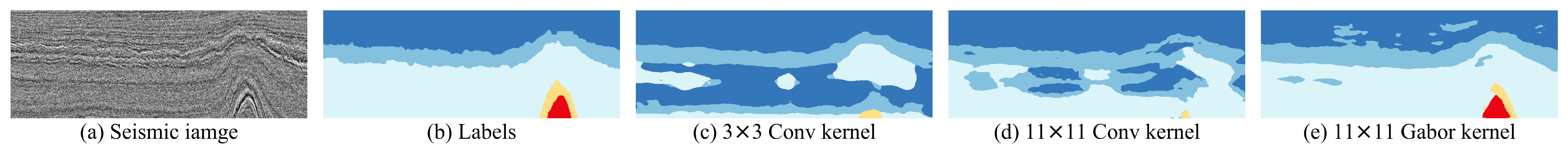}
  \vspace{-12pt}
  \caption{The prediction results for inline \#698 with 30\% salt $\&$ pepper noise using the same CNN models as that in Figures~ \ref{conv3conv11gabor1104test50epochpred55} and \ref{conv3conv11gabor1104test50epochpred309} for different kernels.}
  \vspace{-6pt}
\label{saltpepper395}
\end{figure*}

In this section, we test the robustness of the CNN with different kernels (3$\times$3 Conv kernel, 11$\times$11 Conv kernel, and 11$\times$11 Gabor kernel) on the different types of noise, e.g., salt$\&$pepper noise and speckle noise.
For different kernels, the CNN models used here are the same as that in Figures~ \ref{conv3conv11gabor1104test50epochpred55} and \ref{conv3conv11gabor1104test50epochpred309}.
Figures \ref{saltpepper210} and \ref{saltpepper395} show the predictions on the imaging results with 30\% salt $\&$ pepper noise for inlines \#510 and \#695.
The CNN with the Gabor kernel obviously outperforms others.
Table~\ref{tab:salt-pepper noise} is the quantitative evaluation results on the whole 3D volumes excluding the training profiles.
The CNN with the Gabor kernel shows better performance compared to the CNN with Conv kernel.

\begin{figure*}[!htb]
  \centering
  \includegraphics[width=1\columnwidth]{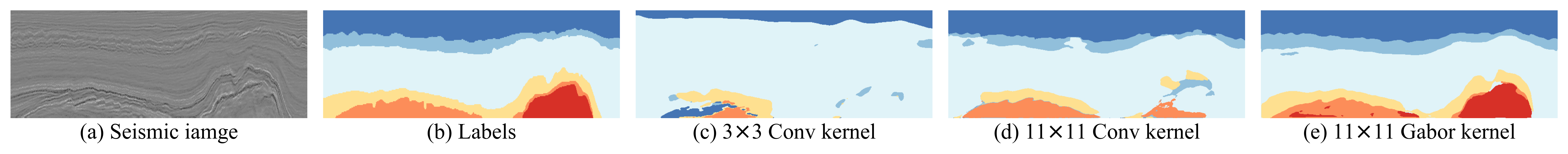}
  \vspace{-12pt}
  \caption{The prediction results for inline \#388 with speckle noise using the same CNN models as that in Figures~ \ref{conv3conv11gabor1104test50epochpred55} and \ref{conv3conv11gabor1104test50epochpred309} for different kernels.}
  \vspace{-6pt}
\label{speckle88}
\end{figure*}

\begin{figure*}[!htb]
  \centering
  \includegraphics[width=1\columnwidth]{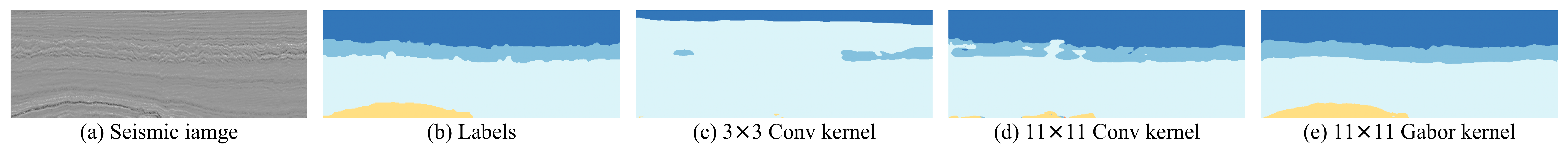}
  \vspace{-12pt}
  \caption{The prediction results for inline \#557 with speckle noise using the same CNN models as that in Figures~ \ref{conv3conv11gabor1104test50epochpred55} and \ref{conv3conv11gabor1104test50epochpred309} for different kernels.}
  \vspace{-6pt}
\label{speckle257}
\end{figure*}

As for the speckle noise, we add the noise by multiplying the mean noise with the image, where the mean and variance of the mean noise are 0 and 0.49.
Then we test the models with different kernels.
The model used here are the same as above.
Figures~\ref{speckle88} and \ref{speckle257} are the predictions for inlines \#388 and \#557.
We also show the quantitative evaluation on the whole 3D volume excluding the training profiles in Table~\ref{tab: speckle noise}.
The metrics demonstrate that the CNN with the Gabor kernel is robust to the noise even dealing with the noise it has never seen during the training (The salt$\&$pepper and speckle noises can be regarded as out-of-distribution noise).

\begin{table*}[!t]
\renewcommand{\arraystretch}{1.25}
    \centering
    \caption{The accuracy of prediction using different kernels with salt\&pepper noise}
    \setlength{\tabcolsep}{0.0125\columnwidth}
    \scalebox{0.8}{
    \begin{tabular}{c|c|c|c|c|c|c|c|c|c}
         \toprule
         Metric &
         \multirow{2}{*}{PA} & \multirow{2}{*}{MCA} &\multirow{2}{*}{FWIU}&
         \multicolumn{6}{c}{Class Accuracy}  \\
         Model & &  &  &  Upper N.S.  & Middle N. S.
         & Lower N.S.& Rijnland/Chalk & Scruff & Zechstein\\         
         \hline
         Conv 3$\times$3& 0.687 & 0.528 & 0.514 & 0.996 & 0.902 & 0.542 & 0.389 & 0.341 & 0.000 \\
         \hline
         Conv 11$\times$11& 0.733 & 0.590 &  0.584 & 0.999 & 0.769 & 0.626 & 0.560 & 0.587 & 0.000 \\
         \hline
         Gabor 11$\times$11
         & \textbf{0.930} & \textbf{0.903} & \textbf{0.879} & 0.966 & 0.978 & 0.919 & 0.854 & 0.738 & 0.961\\
        \bottomrule
    \end{tabular}}
    \label{tab:salt-pepper noise}
\end{table*}

\begin{table*}[!t]
\renewcommand{\arraystretch}{1.25}
    \centering
    \caption{The accuracy of prediction using different kernels with speckle noise}
    \setlength{\tabcolsep}{0.0125\columnwidth}
    \scalebox{0.8}{
    \begin{tabular}{c|c|c|c|c|c|c|c|c|c}
         \toprule
         Metric &
         \multirow{2}{*}{PA} & \multirow{2}{*}{MCA} &\multirow{2}{*}{FWIU}&
         \multicolumn{6}{c}{Class Accuracy}  \\
         Model & &  &  &  Upper N.S.  & Middle N. S.
         & Lower N.S.& Rijnland/Chalk & Scruff & Zechstein\\         
         \hline
         Conv 3$\times$3& 0.619 & 0.302 & 0.408 & 0.354 & 0.164 & 0.998 & 0.181 & 0.114 & 0.000 \\
         \hline
         Conv 11$\times$11& 0.899 & 0.646 &  0.815 & 0.990 & 0.786 & 0.983 & 0.383 & 0.737 & 0.000 \\
         \hline
         Gabor 11$\times$11
         & \textbf{0.954} & \textbf{0.886} & \textbf{0.916} & \textbf{0.997} & \textbf{0.947} & 0.969 & 0.820 & 0.674 & 0.911\\
        \bottomrule
    \end{tabular}}
    \label{tab: speckle noise}
\end{table*}

\section{Discussion}
The Gabor kernel layer added to the UNet CNN architecture in the first layer provides a favorable filter for input seismic images. With its five trainable parameters and functional form, it is inherently biased against noise. The Gabor kernel also allows us to constrain some of the Gabor parameters like wavelength and angle to further control the type of data we admit to the convolutional layers. So if certain angle ranges are preferred in the facies classification, we can emphasize these angles in the network by constraining the range. Though the parameters are still learned as part of the training of the network, the learned values will be within the specified range. The network also allows us to fix some of these parameters if needed. If a certain wavelength is preferred to isolate certain classes of facies, we can set that wavelength in the network and it is not learned.

Though we show the effectiveness of having a Gabor layer for a certain network Unet on a certain task, facies classification, we feel that the Gabor layer for seismic inputs is beneficial for any network, convolutional or not, including networks like ResNet and Vision Transformers. We also speculate that the benefits go beyond facies classification tasks to other seismic tasks like salt segmentation and horizon picking and even denoising. The message here is that Gabor functions are optimal basis functions for seismic data, and having them represent input seismic data should only help in the feature extraction of seismic data.

\section{Conclusions}
Here, we proposed a modified UNet with learnable Gabor convolutional kernels for facies classification. Because Gabor functions are suitable for representing seismic wavefields and images, the features extracted by the Gabor convolutional kernels adapt easily to the seismic texture. Meanwhile, since the Gabor parameters have recognizable meanings with respect to direction and wavelength, the CNN performance can be further improved by constraining these parameters based on our expectations of them in seismic images. The experiments on the Netherland F3 dataset show the effectiveness of the proposed method. The test on salt$\&$pepper and speckle noises further demonstrates the good generalization and robustness of the CNN with Gabor kernels in the first layer.
In many practical exploration applications of deep learning, where we train the CNN model on limited data, adopting Gabor kernels in shallow layers promises high potential in improving the model's generalization.

\section{Appendix}
To evaluate the performance of CNN models with various kernels on the F3 dataset, we use several evaluation metrics. First, we define $G_i$ as the set of pixels that belong to class $i$, and $F_i$ represents the set of pixels predicted as class $i$. Then, the set of correctly predicted pixels is denoted by $G_i \cap F_i$, and the number of elements in a set is extracted by the operator $|\cdot|$. Thus, the evaluation metrics can be defined as follows:
\begin{itemize}
    \item Pixel accuracy (PA) represents the percentage of pixels in the image correctly identified to the right class
\begin{equation}
\mathrm{PA}=\frac{\sum_i\left|F_i \cap G_i\right|}{\sum_i\left|G_i\right|}
\end{equation}
This metric could reflect the overall accuracy of the classification but may fail for the class with limited testing samples.
    \item Class accuracy (CA) represents the prediction accuracy for each class, and $\mathrm{CA}_i$ represents the percentage of pixels correctly classified in a class $i$
\begin{equation}
\mathrm{CA}_i=\frac{\left|F_i \cap G_i\right|}{\left|G_i\right|}
\end{equation}

$MCA$ is defined as the average of CA for all classes
\begin{equation}
\mathrm{MCA}=\frac{1}{n_c} \sum_i \mathrm{CA}_i=\frac{1}{n_c} \sum_i \frac{\left|F_i \cap G_i\right|}{\left|G_i\right|},
\end{equation}
where $n_c$ is the number of classes. The high MAC only happens when the accuracy for each class reaches a high level.

    \item Intersection over union $\left(\mathrm{IU}_i\right)$ is defined as the number of elements of the intersection of $G_i$ and $F_i$ over the number of elements of their union set
\begin{equation}
\mathrm{IU}_i=\frac{\left|F_i \cap G_i\right|}{\left|F_i \cup G_i\right|}
\end{equation}
This metric measures the overlap between the ground truth and the predictions, and higher scores mean better. 
It equals one only in case all pixels were correctly identified to the right class. 
The further metric averaging IU over all classes is defined as the mean intersection over union (Mean IU)
\begin{equation}
\mathrm{Mean\ IU}=\frac{1}{n_c} \sum_i \mathrm{IU}_i=\frac{1}{n_c} \sum_i \frac{\left|F_i \cap G_i\right|}{\left|F_i \cup G_i\right|} .
\end{equation}

To avoid the bias of this metric for the imbalanced classes and sensitivity to the certain class (whose samples are few), the weighted version of the metric results in FWIU:
\begin{equation}
\mathrm{FWIU}=\frac{1}{\sum_i\left|\mathcal{G}_i\right|} \cdot \sum_i\left|\mathcal{G}_i\right| \cdot \frac{\left|\mathcal{F}_i \cup \mathcal{G}_i\right|}{\left|\mathcal{F}_i \cap \mathcal{G}_i\right|} .
\end{equation}
\end{itemize}

\section*{Acknowledgment}
The authors thank KAUST and the DeepWave Consortium sponsors for their support, and Xinquan Huang for useful discussion.
We would also like to thank the SWAG group for the collaborative environment.

\bibliographystyle{plainnat}
\bibliography{reference}

\begin{thebibliography}{26}
\providecommand{\natexlab}[1]{#1}
\providecommand{\url}[1]{\texttt{#1}}
\expandafter\ifx\csname urlstyle\endcsname\relax
  \providecommand{\doi}[1]{doi: #1}\else
  \providecommand{\doi}{doi: \begingroup \urlstyle{rm}\Url}\fi

\bibitem[Alaudah et~al.(2019)Alaudah, Micha{\l}owicz, Alfarraj, and
  AlRegib]{alaudah2019machine}
Yazeed Alaudah, Patrycja Micha{\l}owicz, Motaz Alfarraj, and Ghassan AlRegib.
\newblock A machine-learning benchmark for facies classification.
\newblock \emph{Interpretation}, 7\penalty0 (3):\penalty0 SE175--SE187, 2019.

\bibitem[Ao et~al.(2019)Ao, Li, Zhu, Ali, and Yang]{ao2019identifying}
Yile Ao, Hongqi Li, Liping Zhu, Sikandar Ali, and Zhongguo Yang.
\newblock Identifying channel sand-body from multiple seismic attributes with
  an improved random forest algorithm.
\newblock \emph{Journal of Petroleum Science and Engineering}, 173:\penalty0
  781--792, 2019.

\bibitem[Daugman(1985)]{daugman1985uncertainty}
John~G Daugman.
\newblock Uncertainty relation for resolution in space, spatial frequency, and
  orientation optimized by two-dimensional visual cortical filters.
\newblock \emph{JOSA A}, 2\penalty0 (7):\penalty0 1160--1169, 1985.

\bibitem[Di and AlRegib(2020)]{di2020comparison}
Haibin Di and Ghassan AlRegib.
\newblock A comparison of seismic saltbody interpretation via neural networks
  at sample and pattern levels.
\newblock \emph{Geophysical Prospecting}, 68\penalty0 (2):\penalty0 521--535,
  2020.

\bibitem[Di et~al.(2018)Di, Wang, and AlRegib]{di2018seismic}
Haibin Di, Zhen Wang, and Ghassan AlRegib.
\newblock Seismic fault detection from post-stack amplitude by convolutional
  neural networks.
\newblock In \emph{80th EAGE Conference and Exhibition 2018}, volume 2018,
  pages 1--5. EAGE Publications BV, 2018.

\bibitem[Di et~al.(2020)Di, Li, Maniar, and Abubakar]{di2020seismic}
Haibin Di, Zhun Li, Hiren Maniar, and Aria Abubakar.
\newblock Seismic stratigraphy interpretation by deep convolutional neural
  networks: A semisupervised workflow.
\newblock \emph{Geophysics}, 85\penalty0 (4):\penalty0 WA77--WA86, 2020.

\bibitem[Dumay and Fournier(1988)]{dumay1988multivariate}
Jean Dumay and Frederique Fournier.
\newblock Multivariate statistical analyses applied to seismic facies
  recognition.
\newblock \emph{Geophysics}, 53\penalty0 (9):\penalty0 1151--1159, 1988.

\bibitem[Feng et~al.(2021)Feng, Balling, Grana, Dramsch, and
  Hansen]{feng2021bayesian}
Runhai Feng, Niels Balling, Dario Grana, Jesper~S{\"o}ren Dramsch, and
  Thomas~Mejer Hansen.
\newblock Bayesian convolutional neural networks for seismic facies
  classification.
\newblock \emph{IEEE Transactions on Geoscience and Remote Sensing},
  59\penalty0 (10):\penalty0 8933--8940, 2021.

\bibitem[Krizhevsky et~al.(2017)Krizhevsky, Sutskever, and
  Hinton]{krizhevsky2017imagenet}
Alex Krizhevsky, Ilya Sutskever, and Geoffrey~E Hinton.
\newblock Imagenet classification with deep convolutional neural networks.
\newblock \emph{Communications of the ACM}, 60\penalty0 (6):\penalty0 84--90,
  2017.

\bibitem[Liu et~al.(2020)Liu, Jervis, Li, and Nivlet]{liu2020seismic}
Mingliang Liu, Michael Jervis, Weichang Li, and Philippe Nivlet.
\newblock Seismic facies classification using supervised convolutional neural
  networks and semisupervised generative adversarial networks.
\newblock \emph{Geophysics}, 85\penalty0 (4):\penalty0 O47--O58, 2020.

\bibitem[Lubo-Robles and Marfurt(2019)]{lubo2019independent}
David Lubo-Robles and Kurt~J Marfurt.
\newblock Independent component analysis for reservoir geomorphology and
  unsupervised seismic facies classification in the taranaki basin, new
  zealand.
\newblock \emph{Interpretation}, 7\penalty0 (3):\penalty0 SE19--SE42, 2019.

\bibitem[Roy et~al.(2014)Roy, Romero-Pel{\'a}ez, Kwiatkowski, and
  Marfurt]{roy2014generative}
Atish Roy, Araceli~S Romero-Pel{\'a}ez, Tim~J Kwiatkowski, and Kurt~J Marfurt.
\newblock Generative topographic mapping for seismic facies estimation of a
  carbonate wash, veracruz basin, southern mexico.
\newblock \emph{Interpretation}, 2\penalty0 (1):\penalty0 SA31--SA47, 2014.

\bibitem[Saraswat and Sen(2012)]{saraswat2012artificial}
Puneet Saraswat and Mrinal~K Sen.
\newblock Artificial immune-based self-organizing maps for seismic-facies
  analysis.
\newblock \emph{Geophysics}, 77\penalty0 (4):\penalty0 O45--O53, 2012.

\bibitem[Shi et~al.(2019)Shi, Wu, and Fomel]{shi2019saltseg}
Yunzhi Shi, Xinming Wu, and Sergey Fomel.
\newblock Saltseg: Automatic 3d salt segmentation using a deep convolutional
  neural network.
\newblock \emph{Interpretation}, 7\penalty0 (3):\penalty0 SE113--SE122, 2019.

\bibitem[Tschannen et~al.(2020)Tschannen, Delescluse, Ettrich, and
  Keuper]{tschannen2020extracting}
Valentin Tschannen, Matthias Delescluse, Norman Ettrich, and Janis Keuper.
\newblock Extracting horizon surfaces from 3d seismic data using deep
  learningdeep learning for horizon picking.
\newblock \emph{Geophysics}, 85\penalty0 (3):\penalty0 N17--N26, 2020.

\bibitem[Waldeland and Solberg(2017)]{waldeland2017salt}
AU~Waldeland and AHSS Solberg.
\newblock Salt classification using deep learning.
\newblock In \emph{79th eage conference and exhibition 2017}, volume 2017,
  pages 1--5. EAGE Publications BV, 2017.

\bibitem[Womack and Cruz(1994)]{womack1994seismic}
James~E Womack and Joao~R Cruz.
\newblock Seismic data filtering using a gabor representation.
\newblock \emph{IEEE transactions on geoscience and remote sensing},
  32\penalty0 (2):\penalty0 467--472, 1994.

\bibitem[Wrona et~al.(2018)Wrona, Pan, Gawthorpe, and Fossen]{wrona2018seismic}
Thilo Wrona, Indranil Pan, Robert~L Gawthorpe, and Haakon Fossen.
\newblock Seismic facies analysis using machine learning.
\newblock \emph{Geophysics}, 83\penalty0 (5):\penalty0 O83--O95, 2018.

\bibitem[Wu et~al.(2019{\natexlab{a}})Wu, Zhang, Lin, Cao, and
  Lou]{wu2019semiautomated}
Hao Wu, Bo~Zhang, Tengfei Lin, Danping Cao, and Yihuai Lou.
\newblock Semiautomated seismic horizon interpretation using the
  encoder-decoder convolutional neural network.
\newblock \emph{Geophysics}, 84\penalty0 (6):\penalty0 B403--B417,
  2019{\natexlab{a}}.

\bibitem[Wu et~al.(2019{\natexlab{b}})Wu, Liang, Shi, and
  Fomel]{wu2019faultseg3d}
Xinming Wu, Luming Liang, Yunzhi Shi, and Sergey Fomel.
\newblock Faultseg3d: Using synthetic data sets to train an end-to-end
  convolutional neural network for 3d seismic fault segmentation: Geophysics,
  84.
\newblock \emph{IM35--IM45}, 2019{\natexlab{b}}.

\bibitem[Wu et~al.(2019{\natexlab{c}})Wu, Shi, Fomel, Liang, Zhang, and
  Yusifov]{wu2019faultnet3d}
Xinming Wu, Yunzhi Shi, Sergey Fomel, Luming Liang, Qie Zhang, and Anar~Z
  Yusifov.
\newblock Faultnet3d: Predicting fault probabilities, strikes, and dips with a
  single convolutional neural network.
\newblock \emph{IEEE Transactions on Geoscience and Remote Sensing},
  57\penalty0 (11):\penalty0 9138--9155, 2019{\natexlab{c}}.

\bibitem[Zhang et~al.(2015)Zhang, Zhao, Jin, and Marfurt]{zhang2015brittleness}
Bo~Zhang, Tao Zhao, Xiaochun Jin, and Kurt~J Marfurt.
\newblock Brittleness evaluation of resource plays by integrating petrophysical
  and seismic data analysis.
\newblock \emph{Interpretation}, 3\penalty0 (2):\penalty0 T81--T92, 2015.

\bibitem[Zhao(2018)]{zhao2018seismic}
Tao Zhao.
\newblock Seismic facies classification using different deep convolutional
  neural networks.
\newblock In \emph{2018 SEG International Exposition and Annual Meeting}.
  OnePetro, 2018.

\bibitem[Zhao and Mukhopadhyay(2018)]{zhao2018fault}
Tao Zhao and Pradip Mukhopadhyay.
\newblock A fault detection workflow using deep learning and image processing.
\newblock In \emph{2018 SEG international exposition and annual meeting}.
  OnePetro, 2018.

\bibitem[Zhao et~al.(2014)Zhao, Jayaram, Marfurt, and
  Zhou]{zhao2014lithofacies}
Tao Zhao, Vikram Jayaram, Kurt~J Marfurt, and Huailai Zhou.
\newblock Lithofacies classification in barnett shale using proximal support
  vector machines.
\newblock In \emph{SEG Technical Program Expanded Abstracts 2014}, pages
  1491--1495. Society of Exploration Geophysicists, 2014.

\bibitem[Zhao et~al.(2017)Zhao, Li, and Marfurt]{zhao2017constraining}
Tao Zhao, Fangyu Li, and Kurt~J Marfurt.
\newblock Constraining self-organizing map facies analysis with stratigraphy:
  An approach to increase the credibility in automatic seismic facies
  classification.
\newblock \emph{Interpretation}, 5\penalty0 (2):\penalty0 T163--T171, 2017.

\end{thebibliography}
\end{document}